\documentclass{SciPost}

\hypersetup{
    colorlinks,
    linkcolor={red!50!black},
    citecolor={blue!50!black},
    urlcolor={blue!80!black}
}

\usepackage[bitstream-charter]{mathdesign}
\DeclareSymbolFont{usualmathcal}{OMS}{cmsy}{m}{n}
\DeclareSymbolFontAlphabet{\mathcal}{usualmathcal}

\usepackage{mathtools}   
\usepackage{bm}
\usepackage{dsfont}      
\providecommand{\ud}{}\renewcommand{\ud}{\mathrm{d}}
\providecommand{\e}{}\renewcommand{\e}{\mathrm{e}}

\fancypagestyle{SPstyle}{
\fancyhf{}
\lhead{\colorbox{scipostblue}{\bf \color{white} ~SciPost Physics Core }}
\rhead{{\bf \color{scipostdeepblue} ~Submission }}

\fancyfoot[C]{\textbf{\thepage}}
}

\begin{document}

\pagestyle{SPstyle}

\begin{center}{\Large \textbf{\color{scipostdeepblue}{
Sound waves in a backreaction-affected analogue spacetime: acoustic metric, mass, and correlations in Bose-Einstein condensates
}}}\end{center}

\begin{center}\textbf{
Sang-Shin Baak\textsuperscript{1} and
Hyun-Yong Lee\textsuperscript{2,3,4$\star$}
}\end{center}

\begin{center}
{\bf 1} Natural Science Research Institute, Korea University, Sejong 30019, South Korea
\\
{\bf 2} Division of Semiconductor Physics, Korea University, Sejong 30019, South Korea
\\
{\bf 3} Department of Applied Physics, Graduate School, Korea University, Sejong 30019, South Korea
\\
{\bf 4} Center for Extreme Quantum Matter and Functionality, Sungkyunkwan University, Suwon 16419, South Korea
\\[\baselineskip]
$\star$ \href{mailto:hyunyong@korea.ac.kr}{\small hyunyong@korea.ac.kr}
\end{center}

\section*{\color{scipostdeepblue}{Abstract}}
\textbf{\boldmath{%
Analogue gravity treats sound as a field on an effective spacetime set by the background flow, but quantum fluctuations feed back on this background. Whether a condensate modified by such backreaction still defines a spacetime for the quantum field remains open. Using a number-conserving expansion of a Bose-Einstein condensate, we show that in the sound-wave and eikonal approximations the first-order phase fluctuation propagates on an acoustic metric of the backreaction-corrected condensate and acquires a spacetime-dependent mass that cannot be absorbed into the metric. For a finite, quasi-one-dimensional gas after an interaction quench, backreaction enhances the equal-point phase correlation by about $10\%$ in the bulk and up to $20\%$ near the walls while the depletion remains below $2\%$; the bulk enhancement survives finite imaging resolution.
}}

\vspace{\baselineskip}


\vspace{10pt}
\noindent\rule{\textwidth}{1pt}
\tableofcontents
\noindent\rule{\textwidth}{1pt}
\vspace{10pt}

\section{Introduction}\label{sec:intro}
Sound waves in an inviscid, barotropic, and irrotational fluid obey the equation of a massless scalar field on an effective curved spacetime \cite{Unruh1981}. This correspondence underlies analogue gravity \cite{Barcelo2011}, including laboratory observations of analogue Hawking radiation \cite{Philbin2008,Silke2011,Steinhauer2016,Steinhauer2019} and quantum simulations of fields in curved spacetime \cite{Tajik2025,Schutzhold2025}. Going beyond a prescribed background requires accounting for backreaction: quantum fluctuations modify the medium, which in turn changes their propagation. In gravity this feedback involves a renormalized stress-energy expectation value and a self-consistent geometry \cite{Hu2008}. Bose-Einstein condensates offer a microscopic setting in which the background and the quantum field belong to the same many-body system.

Effects of backreaction have been measured in analogue black holes with water waves \cite{Patrick2021} and light \cite{Procopio2026}, and backreaction has been studied in acoustic black holes \cite{Balbinot2005PRD,Balbinot2005PRL,Schuetzhold2008JPA,Tricella2020,Balbinot2025,Ciliberto2025}, analogue preheating \cite{Butera2023,Butera2023prl}, parametric oscillations \cite{Butera2019}, nonlinear fluids \cite{Datta2022} and optics \cite{Baak2024}, and Bose gases beyond the Bogoliubov approximation \cite{Fischer2005,Fischer2007,Baak2022,Ribeiro2025}. For analogue gravity, however, a change of the condensate is not by itself a change of spacetime. At leading order the condensate defines the acoustic spacetime on which the linear fluctuation propagates, but a condensate modified by backreaction need not define a spacetime for that fluctuation; unless it does, the correction describes feedback within a many-body system rather than backreaction on an analogue spacetime. Conserving gapless mean-field theory yields an acoustic metric with backreaction, but without separating the condensate from its fluctuations, so that the metric depends directly on anomalous averages of the fluctuation field \cite{Kurita2010}. Amplitude-phase \cite{Balbinot2025,Ciliberto2025} and number-conserving \cite{Fischer2005,Fischer2007,Baak2022,Baak2024,Ribeiro2025} expansions separate the condensate, its quantum fluctuation, and the resulting background correction order by order, but do not address whether the corrected condensate still acts as a metric. Beyond linear order, a quantum metric that contains the first-order fluctuation has been derived for second-order perturbations from the Madelung fluid equations \cite{Pal2024}; this geometry is seen by a different field, the second-order perturbation, and describes its propagation on the condensate and the first-order fluctuation rather than the feedback of the first-order fluctuation on its own spacetime. Whether backreaction in these systems is a backreaction of analogue gravity, i.e., a change of the spacetime seen by the quantum field itself, has therefore remained open.

Here we address this question for the first-order phase field, using a number-conserving expansion \cite{Girardeau1959,Gardiner1997,Girardeau1998,Castin1998,Lieb2000,Gardiner2007} that organizes the feedback in powers of $1/N$, with $N$ the total particle number. This also connects corrections to sound propagation with correlations calculated from the same coupled microscopic dynamics.

We show that, in the sound-wave and eikonal approximations, this backreaction-affected phase field acquires a spacetime-dependent effective-mass term generated by the sources in the modified continuity and Euler equations. In the geometric formulation, the acoustic metric retains its usual functional form, with corrected sound velocity, density, and flow velocity. The mass term appears in addition to these metric corrections. Both effects enter at order $1/N$; neglecting them recovers the standard massless sound equation. Backreaction in a Bose-Einstein condensate is therefore a change of the analogue spacetime seen by the quantum field, supplemented by a mass term that has no geometric counterpart.

As an example, we solve the coupled condensate and Bogoliubov equations for a homogeneous, finite-size, quasi-one-dimensional Bose gas quenched from a noninteracting state at $t=0$. At $t=5$, while depletion remains below $2\%$, backreaction enhances the canonically normal-ordered equal-point phase correlation by about $10\%$ in the bulk and up to about $20\%$ within a few healing lengths of the edges. Correlations between distant points remain small. The bulk enhancement survives the finite-resolution imaging model considered here, suggesting a route to probing feedback through quantum correlations. Experimental access additionally requires accounting for thermal fluctuations and the contact contribution, together with a suitably high density and a tuned interaction strength.

The paper is organized as follows. In section~\ref{sec:NCE}, we review the number-conserving expansion and derive the continuity and Euler equations of the classical fluid modified by backreaction. In section~\ref{sec:sound}, we derive the equation of motion for the phase fluctuation and show that it takes the form of a massive Klein-Gordon equation on an acoustic spacetime. In section~\ref{sec:model}, we introduce the model, and in section~\ref{sec:results} we present the numerical results and discuss the conditions for observing them. Section~\ref{sec:conclusion} concludes. Details of the derivation, of the model, and of the numerical method, the validity regime of the expansion, and the effect of finite imaging resolution are given in the appendices.

\section{Number-conserving expansion and modified classical fluid dynamics}\label{sec:NCE}
The Hamiltonian for the field operator describing Bose particles in $d$-dimensional spacetime in the s-wave approximation is
\begin{align}\label{eq:BoseHamiltonian}
    \hat{H} = \int \ud^{{d-1}} x \bigg[&\frac{1}{2}\nabla\hat{\Psi}^\dagger\cdot\nabla\hat{\Psi} + U_{\rm ext}\hat{\Psi}^\dagger\hat{\Psi}+ \frac{g}{2}\hat{\Psi}^\dagger\hat{\Psi}^\dagger\hat{\Psi}\hat{\Psi}\bigg],
\end{align}
where $U_{\rm ext}=U_{\rm ext}(t,\vec{x})$ is an external potential, and $g$ is a coupling constant. We set $m=\hbar=1$ for simplicity. Since backreaction is at least a second-order effect, an expansion beyond linear order is needed \cite{Braunstein2023}. The number-conserving expansion provides a systematic way to go beyond linear order \cite{Girardeau1959,Gardiner1997,Girardeau1998,Castin1998,Lieb2000,Gardiner2007,Billam2012,Billam2013}.
Adopting the number-conserving mean-field ansatz
\begin{equation*}
    \hat{\Psi} = \phi_c(1 + \hat{\psi})\frac{\hat{A}}{\sqrt{\hat{N}}},
\end{equation*}
where $\hat{N} = \hat{A}^\dagger\hat{A}$ is the total particle-number operator, $\phi_c=\mathcal{O}(\sqrt{N})$ is the $U(1)$-conserving order parameter, and $\hat{\psi}=\mathcal{O}(N^{-1/2})$ is the $U(1)$-conserving quantum fluctuation on it. Moreover, if $g=\mathcal{O}(N^{-1})$ we can expand Eq.~\eqref{eq:BoseHamiltonian} in powers of $1/N$ and obtain the amended Gross-Pitaevskii equation \cite{Fischer2005,Ralf2006}:
\begin{align}\label{eq:amendedGPpsi}
    i\partial_t\phi_c &= \Big(-\frac{1}{2}\nabla^2+U_{\rm ext} +g|\phi_c|^2\Big)\phi_c +g|\phi_c|^2\big(2\langle\hat{\psi}^\dagger\hat{\psi}\rangle + \langle\hat{\psi}^2 \rangle\big)\phi_c,
\end{align}
and the Bogoliubov-de Gennes equation:
\begin{align}\label{eq:BdGpsi}
    i\partial_t\hat{\psi} &= \Big(-\frac{1}{2}\nabla^2-\frac{\nabla\phi_c}{\phi_c}\cdot\nabla\Big)\hat{\psi} +g|\phi_c|^2(\hat{\psi}+\hat{\psi}^\dagger) -g|\phi_c|^2\big(2\langle\hat{\psi}^\dagger\hat{\psi}\rangle +\langle\hat{\psi}^2\rangle\big)\hat{\psi}.
\end{align}
Equation~\eqref{eq:BdGpsi} is the Bogoliubov-de Gennes equation for the fluctuation of the field operator on the backreaction-corrected condensate \cite{Fischer2005,Fischer2007}, written for the relative fluctuation $\hat{\psi}$; its last term arises from the time dependence of $\phi_c$. Note that the last term of both equations is $\mathcal{O}(1/N)$ compared to the other terms.

Now, we will derive the equation of motion for the classical fluid together with the quantum backreaction correction. Let us define the classical density and current by
\begin{equation}\label{eq:classicalobservables}
    \rho_c:=\phi_c^*\phi_c,\qquad \vec{\jmath}_c := \Im[\phi_c^*\nabla\phi_c].
\end{equation}
From Eq.~\eqref{eq:amendedGPpsi}, we obtain a continuity-like equation:
\begin{equation}\label{eq:ClassCont}
    \partial_t\rho_c + \nabla\cdot(\vec{\jmath}_c) = \rho_c\Delta_{\rm C},
\end{equation}
where
\begin{equation*}
    \Delta_{\rm C} \coloneq ig\rho_c(\langle\hat{\psi}^{\dagger2}\rangle-\langle\hat{\psi}^2\rangle)
\end{equation*}
describes the modification of the continuity equation for the classical field due to the quantum fluctuations.
Let us put the Madelung representation, $\phi_c \equiv \sqrt{\rho_c}\e^{i\theta_c}$, into Eq.~\eqref{eq:amendedGPpsi}. Then we obtain an Euler-like equation:
\begin{equation}\label{eq:ClassEuler}
    \partial_t\theta_c-\frac{1}{2\sqrt{\rho_c}}\nabla^2\sqrt{\rho_c}+\frac{1}{2} (\nabla \theta_c)^2 + U_{\rm ext} +g\rho_c + \Delta_{\rm E} = 0,
\end{equation}
where
\begin{equation*}
    \Delta_{\rm E} \coloneq g\rho_c\Big(2\langle\hat{\psi}^\dagger\hat{\psi}\rangle + \frac{1}{2}(\langle\hat{\psi}^2\rangle+\langle\hat{\psi}^{\dagger2}\rangle)\Big)
\end{equation*}
describes the modification of the Euler equation for the classical field by the quantum fluctuations.

Note that $\Delta_{\rm C}=\mathcal{O}(1/N)$ and $\Delta_{\rm E}=\mathcal{O}(1/N)$ are determined solely by the classical fluid variables, through Eqs.~\eqref{eq:ClassCont} and \eqref{eq:ClassEuler}, respectively. Neglecting these terms recovers the usual Bogoliubov-de Gennes equation; however, since they constitute the leading-order, $\mathcal{O}(1/N)$, correction from backreaction, it is essential to retain them in order to capture its effect.

\section{Quantum fluid and sound wave}\label{sec:sound}
Now, let us write the number-conserving ansatz in the Madelung representation,
\begin{equation*}
    \hat{\Psi} = \sqrt{\rho_c}\e^{i\theta_c}\Big(1+\frac{\delta\hat{\rho}}{2\rho_c}+i\delta\hat{\theta}\Big)\frac{\hat{A}}{\sqrt{\hat{N}}}.
\end{equation*}
In the usual Bogoliubov (linear) regime, this is the standard route to the analogue spacetime. Without loss of generality, we can take $\delta\hat{\rho}$ and $\delta\hat{\theta}$ to be Hermitian:
\begin{align*}
    \delta\hat{\rho} &\coloneq 2\rho_c\Re[\hat{\psi}] = \rho_c(\hat{\psi} + \hat{\psi}^\dagger),\\ \delta\hat{\theta} &\coloneq \Im[\hat{\psi}] = \frac{1}{2i}(\hat{\psi}-\hat{\psi}^\dagger).
\end{align*}
From the Bogoliubov-de Gennes equation \eqref{eq:BdGpsi} and using the continuity-like equation \eqref{eq:ClassCont}, we obtain (see appendix~\ref{app:derivation}):
\begin{align}
    \Big(\partial_t +\vec{v}_c\cdot\nabla+ \frac{1}{2}\Delta_{\rm C}\Big)\delta\hat{\theta} &= -D\Big(\frac{\delta\hat{\rho}}{\rho_c}\Big)\label{eq:QuantumFluid2}\\
    \frac{1}{\rho_c}\Big(\partial_t +\nabla\cdot\vec{v}_c - \frac{1}{2}\Delta_{\rm C}\Big)\delta\hat{\rho}&= -\Big(\frac{1}{\rho_c}\nabla\cdot(\rho_c\nabla) +2\Delta_{\rm E}\Big)\delta\hat{\theta}\label{eq:QuantumFluid3}
\end{align}
where in Eq.~\eqref{eq:QuantumFluid3}, $\nabla\cdot \vec{v}_cf = \nabla\cdot(\vec{v}_cf)$, and in Eq.~\eqref{eq:QuantumFluid2} $D$ is an operator defined as
\begin{equation*}
    D \coloneq g\rho_c -\frac{1}{4\rho_c}\nabla\cdot(\rho_c\nabla)-\frac{1}{2}\Delta_{\rm E}.
\end{equation*}
In analogue gravity, we are interested in the sound wave approximation or eikonal approximation. In both cases, the differential operator in $D$ can be neglected or replaced by the wave number of quantum field; in the latter case the metric depends on the wave number (rainbow geometry) \cite{Weinfurtner2009}. From now on, we assume that $D$ is replaced by one of these approximations. Let us define the generalized sound velocity as $c^2=D$. Then the equation of motion for $\delta\hat{\theta}$ can be written in the form
\begin{equation}\label{eq:App:GeometricProto}
    \partial_\mu(f^{\mu\nu}\partial_{\nu}\delta\hat{\theta}) + V \delta\hat{\theta} = 0,
\end{equation}
where
{\renewcommand{\arraystretch}{1.5}\begin{equation*}
    f^{\mu\nu} = \frac{\rho_c}{c^2}\left(\begin{array}{{@{}c|c@{}}}
        -1 &-v_c^j\\
        \hline
        -v_c^i & c^2\delta^{ij} - v_c^iv_c^j
    \end{array}\right),
\end{equation*}}
and using continuity-like equation Eq.~\eqref{eq:ClassCont}, we obtain
\begin{align*}
    V = -\frac{\rho_c}{2c^2}\Big(\partial_t+{\vec{v}_c\cdot\nabla}+\frac{1}{2}\Delta_{\rm C}\Big)\Delta_{\rm C}+2\rho_c\Delta_{\rm E}.
\end{align*}
For geometric formulation, let us identify
\begin{equation*}
    f^{\mu\nu} = \sqrt{-\mathfrak{g}}\,\mathfrak{g}^{\mu\nu}
\end{equation*}
then we obtain
{\renewcommand{\arraystretch}{1.5}\begin{equation*}
    \mathfrak{g}_{\mu\nu} = \Big(\frac{\rho_c}{c}\Big)^{\frac{2}{{d-2}}}\left(\begin{array}{{@{}c|c@{}}}
        -(c^2-v_c^2) &-v_c^j \ \ \\
        \hline
        -v_c^i & \delta^{ij}
    \end{array}\right),
\end{equation*}}
and finally,\footnote{For $d=2$ this expression is ill-defined. This is a formal rather than a fundamental problem, which can be resolved by adding two extra dimensions \cite{Barcelo2004}.} Eq.~\eqref{eq:App:GeometricProto} can be written in the form of massive Klein-Gordon equation:
\begin{equation*}
    \Big(\frac{1}{\sqrt{-\mathfrak{g}}}\partial_\mu(\sqrt{-\mathfrak{g}}\mathfrak{g}^{\mu\nu}\partial_\nu) - m^2\Big)\delta\hat{\theta}=0,
\end{equation*}
where
\begin{align*}
    m^2 &\coloneq -\frac{1}{\sqrt{-\mathfrak{g}}}V(t,\vec{x})=\Big(\frac{c^2}{\rho_c^{{2}}}\Big)^{\frac{1}{d-2}}\Big[\frac{1}{2c^2}\Big(\partial_t+{\vec{v}_c\cdot\nabla}+\frac{1}{2}\Delta_{\rm C}\Big)\Delta_{\rm C}-2\Delta_{\rm E}\Big]
\end{align*}
is the spacetime-dependent squared mass.

Since $\Delta_{\rm C}$ and $\Delta_{\rm E}$ are determined by the classical fluid variables, the metric and the mass can be written in terms of the classical fluid alone. The mass term arises from the same backreaction terms that modify the continuity and Euler equations, and thus captures the part of the backreaction that is not described by the change of the metric; it enters the wave equation as an additional term beyond the effective metric. A spacetime-dependent effective mass of a quantum field is also familiar in cosmology, for example in preheating, where matter fields acquire a time-dependent effective mass through their coupling to the oscillating inflaton, whose dynamics is in turn modified by the backreaction of the produced particles \cite{Kofman1994,Kofman1997,Traschen1990}.

The two terms in $m^2$ have different origins. The first, proportional to $\big(\partial_t+\vec{v}_c\cdot\nabla+\frac{1}{2}\Delta_{\rm C}\big)\Delta_{\rm C}$, originates from the exchange of particles between the condensate and its fluctuations and vanishes when the condensate number is conserved, whereas the second, $-2\Delta_{\rm E}$, originates from the shift of the chemical potential of the condensate and depends on how the fluctuations are treated beyond leading order (appendix~\ref{app:validity}). A detailed analysis of the two contributions, including other treatments of the fluctuations, is left for future work.

\section{The model}\label{sec:model}
In the previous section, we showed that the sound wave in the backreaction-affected dynamical spacetime obeys a massive Klein-Gordon equation. Moreover, the backreaction effect appears as an $\mathcal{O}(1/N)$ deviation from the massless Klein-Gordon equation derived without backreaction. Correlation functions of the phase field and its gradient are now measured in cold-atom quantum simulators \cite{Tajik2023,Murtadho2025}. Motivated by this, in this section we use a finite-size homogeneous one-dimensional Bose gas as an example and compare the correlation function of the sound wave in the linear theory without backreaction with that in the dynamical background with backreaction. In the example below, the leading-order condensate is homogeneous and stationary, but the backreaction sources $\Delta_{\rm C}$ and $\Delta_{\rm E}$ evolve in time after the quench and vary in space near the walls (appendix~\ref{app:validity}), so that the mass is nonzero both in the bulk and near the walls. The correlation function is computed from the full Bogoliubov dynamics, which contain the same backreaction sources that generate the metric corrections and the mass in the sound-wave and eikonal approximations; it therefore reflects their combined effect. Since the effect of backreaction accumulates as the system evolves, the difference between the correlation functions is expected to become more significant with time, as long as the expansion remains valid. The experimental conditions required to observe this difference are discussed in section~\ref{sec:experiment}.

We consider a simple one-dimensional model with external potential
\begin{equation*}
U_{\rm ext}=\mu+\frac{1}{2}\partial_x[\delta(x-\ell/2)-\delta(x+\ell/2)],
\end{equation*}
where $\mu$ is the chemical potential. This model is analysed in detail in Ref.~\cite{Baak2022}. For this model, we use units such that $c_0=\sqrt{g\rho_0}=1$ where $\rho_0$ is the leading order condensate density obtained from the Gross-Pitaevskii equation. Choosing these units, spatial coordinates are expressed in terms of the healing length $\xi_0=1/\sqrt{g\rho_0}=1$, and time is expressed in units of $\xi_0^2$.

Since a stationary state that includes backreaction is not known a priori, we prepare the system in the noninteracting ground state and quench the interaction at $t=0$, which provides a well-defined initial condition. Moreover, because the Bogoliubov-de Gennes equation is linear and $\sigma_3$-pseudoHermitian, we can do the mode expansion at $t=0$,
\begin{equation*}
    \hat{\psi}(t,x)=\sum_{n=0}^{\infty}\left[\hat{a}_{n}u_{n}(t,x)+\hat{a}^{\dagger}_{n}v_{n}^{*}(t,x)\right].
\end{equation*}
The correlation function at time $t$ of the sound wave is
\begin{align*}
    C_{\delta\theta}(t,x,y)\coloneq\langle\delta\hat{\theta}(t,x)\delta\hat{\theta}(t,y)\rangle,
\end{align*}
where $\langle \ \cdot \ \rangle\coloneq \langle 0|\ \cdot \ |0\rangle$. The vacuum state is defined such that $\forall n\in \mathbb{N}: \hat{a}_n|0\rangle = 0$. In the numerical calculation, we truncate the mode sum at $n_{\max}=1000$. Note that since the background is dynamical, subtracting $\delta(x-y)/\rho_0$ does not necessarily yield a finite value. Since we are working in the Heisenberg picture and we know the time evolution of the mode functions, instead, we can do the normal ordering procedure. We obtain
\begin{align*}
    C_{\delta\theta} = \frac{1}{4}\Pi + G_{\delta\theta},
\end{align*}
where
\begin{align*}
    \Pi(x,y)&\coloneq\langle[\hat{\psi}(t,x),\hat{\psi}^\dagger(t,y)]\rangle
\end{align*}
is the subtraction term coming from the \emph{canonical} normal ordering. The subtraction term yields
\begin{align*}
     G_{\delta\theta}(t,x,y) = \frac{1}{4}\sum_n\big[2v_n(x)v_n^*(y)-u_n(x)v_n^*(y)-v_n(x)u_n^*(y)\big].
\end{align*}
Note that if we use the mode functions in the linear theory, we obtain the correlation function, subtraction term, and normal ordered correlation function for the linear theory. The explicit mode functions are given in appendix~\ref{app:model}, and the numerical method in appendix~\ref{app:numerics}.

\section{Results}\label{sec:results}
\begin{figure}[tb]
\centering
\includegraphics[width=0.7\textwidth]{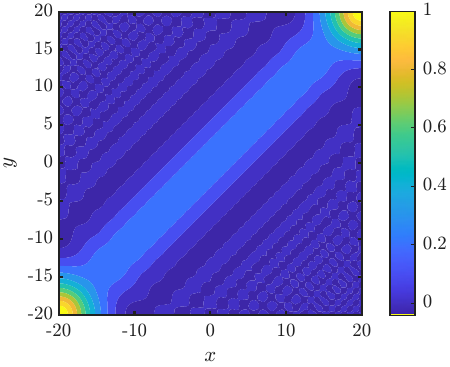}
\caption{Truncated difference between correlation function with and without backreaction, $\rho_0\big(G_{\delta\theta}(x,y)-G_{\delta\theta_0}(x,y)\big)$ at $t=5$ with the lowest 1000 modes. Here and in the following plots, for the numerical simulation, we choose $g\rho_0 = 1$ with $\rho_0=150$. Note that the correlation function is $\mathcal{O}(1/N)$. Therefore, we rescale the correlation with $\rho_0$. Lengths are in units of $\xi_0$ and times in units of $\xi_0^2$.}
\label{fig:GDiff}
\end{figure}
\begin{figure}[tb]
\centering
\includegraphics[width=0.7\textwidth]{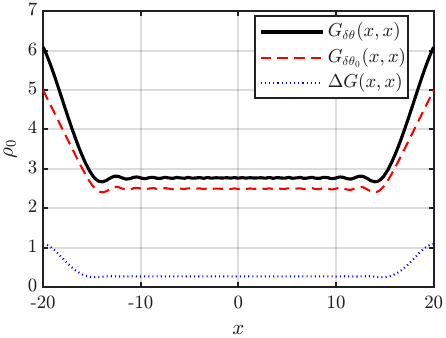}
\caption{Truncated, rescaled correlation function at the same point $G_{\delta\theta}(x,x)$ (black solid line) and $G_{\delta\theta_0}(x,x)$ (red dashed line) and their difference $\Delta G(x,x) = G_{\delta\theta}(x,x)-G_{\delta\theta_0}(x,x)$ (blue dotted line) at $t=5$ with 1000 modes. Note that at the boundary the correlation is enhanced by about 20\%.}
\label{fig:GSame}
\end{figure}
\begin{figure}[tb]
\centering
\includegraphics[width=0.7\textwidth]{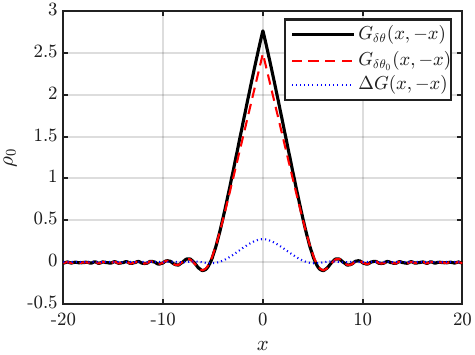}
\caption{Truncated, rescaled correlation function at the opposite point $G_{\delta\theta}(x,-x)$ (black solid line) and $G_{\delta\theta_0}(x,-x)$ (red dashed line) and their difference $\Delta G(x,-x) = G_{\delta\theta}(x,-x)-G_{\delta\theta_0}(x,-x)$ (blue dotted line) at $t=5$ with 1000 modes.}
\label{fig:GOpposite}
\end{figure}
\begin{sloppypar}
Figure~\ref{fig:GDiff} shows the two-dimensional contour plot of the truncated difference between the canonically normal-ordered correlation functions with and without backreaction, $\rho_0\big(G_{\delta\theta}(x,y)-\allowbreak G_{\delta\theta_0}(x,y)\big)$, at $t=5$ with the lowest $1000$ modes. The correlation is visibly enhanced by backreaction, and the pattern of the difference resembles that of the correlation function itself. To interpret this enhancement more clearly, we plot the correlation functions separately along the lines $y=x$ and $y=-x$.
\end{sloppypar}

Since we subtract the canonical contact term, the correlation function at the same point converges with the mode cutoff, with a change of $0.02\%$ for the correlation function and $0.5\%$ for its backreaction contribution between $n_{\max}=500$ and $1000$ (appendix~\ref{app:numerics}). Figure~\ref{fig:GSame} shows the correlation function at the same point. Backreaction increases the correlation by about $10\%$ in the bulk of the condensate. Near the edges the increase is larger: it reaches about $20\%$ at the wall and decays to the bulk value within about six healing lengths. We do not attempt to separate the contribution of the mass from that of the metric corrections; the enhancement should be understood as the combined effect of the backreaction sources on the propagation of the phase field.

Figure~\ref{fig:GOpposite} shows the canonically normal-ordered correlation functions $\rho_0G_{\delta\theta}(x,-x)$ and $\rho_0G_{\delta\theta_0}(x,-x)$ between opposite parts of the condensate. When the two points are far apart, the correlation itself is small, and so is the difference, which makes it hard to discern between the cases with and without backreaction.

\subsection{Conditions for observation}\label{sec:experiment}
We comment on the conditions under which the effect could be observed. Since the results depend only on $\rho_0\xi_0$, $\ell/\xi_0$, and $t$, they can be mapped to different physical parameters by choosing $\xi_0$. For definiteness we consider a box of $46\,\mu$m, comparable to the boxes of present experiments \cite{Murtadho2025,Moller2021}, which requires $\xi_0\approx1.15\,\mu$m, a linear density of about $130\,\mu\mathrm{m}^{-1}$ per gas (about $6000$ atoms), and a sound velocity of $0.64\,\mu$m/ms, so that $t=5$ corresponds to about $9$ ms. The interaction strength is then about twenty times weaker than for $^{87}$Rb on present atom chips, which calls for atomic species or magnetic fields with tunable interactions \cite{Chin2010}. In this setting a camera pixel of $2\,\mu$m corresponds to $1.7\,\xi_0$ and a point spread function of width $3\,\mu$m to $2.6\,\xi_0$; other box lengths are considered in appendix~\ref{app:imaging}. After convolution with such a Gaussian and averaging over the pixels, the relative increase of the same-point phase correlation in the bulk, with the unobservable global phase mode removed, remains about $10\%$ for widths up to $9\,\mu$m, and the relative increase of the velocity correlation between neighboring pixels is of the same order, about $10$ to $20\%$. The stronger enhancement near the walls is still visible in the outermost pixel (about $20\%$ for a width of $3\,\mu$m), but it requires a phase measurement that extends to the edges of the gas. Further conditions have to be met for a quantitative comparison. Our calculation is performed at zero temperature and with canonical normal ordering, whereas measured correlations also contain classical thermal phase fluctuations and the subtracted term; the inclusion of a thermal initial state is left for future work. Moreover, the number-conserving expansion requires a small depletion, which restricts the interaction strength and the evolution time to the regime in which backreaction remains a small correction. For the parameters used here the depletion stays below $2\%$ (appendix~\ref{app:validity}). The limitations of the scheme used for the fluctuations are discussed in appendix~\ref{app:validity}.

\section{Conclusion}\label{sec:conclusion}
We have studied whether the quantum field still propagates on an analogue spacetime once backreaction is included, a question that complements existing studies of how backreaction modifies the classical background. Using the number-conserving expansion, we showed that the sound wave in the backreaction-affected dynamical spacetime obeys a massive Klein-Gordon equation with a spacetime-dependent mass. The effective metric keeps the form of the case without backreaction, with the sound velocity, fluid density, and fluid velocity corrected at order $1/N$, and the mass term arises at the same order. Backreaction in a Bose-Einstein condensate is therefore a change of the analogue spacetime seen by the quantum field, supplemented by a mass term that has no geometric counterpart, and the quantum field on a backreaction-affected spacetime can be studied within the analogue-spacetime framework.

As an example, we considered a homogeneous, finite-size, quasi-one-dimensional Bose gas and calculated the mode functions and the correlation function. Backreaction increases the normal-ordered correlation function at the same point, by about $10\%$ in the bulk and by up to about $20\%$ near the edges at $t=5$, whereas for distant points the correlation itself is small and the difference is hard to discern. The bulk increase persists after accounting for a finite imaging resolution of a few micrometers. We comment further on how the particle number, the interaction strength, and the temperature required here could be approached in future experiments in appendix~\ref{app:outlook}.

\section*{Acknowledgements}

\paragraph{Funding information}
H.-Y.L. and S.-S.B. were supported by the Basic Science Research Program through the National Research Foundation of Korea funded by the Ministry of Science and ICT [Grant No. RS-2025-16064392].

\paragraph{Data availability}
The data that support the findings of this study are available from the authors upon reasonable request.

\begin{appendix}
\numberwithin{equation}{section}

\section{Details of the derivation}\label{app:derivation}
In this appendix, we give the intermediate steps of the derivation in sections~\ref{sec:NCE} and \ref{sec:sound}. Using
\begin{equation*}
    \Delta_{\rm E}+\frac{i}{2}\Delta_{\rm C} = g\rho_c\Big(2\langle\hat{\psi}^\dagger\hat{\psi}\rangle + {\langle\hat{\psi}^{2}\rangle}\Big),
\end{equation*}
the Bogoliubov-de Gennes equation \eqref{eq:BdGpsi} can be written in the form
\begin{align}\label{eq:BdGpsifluid}
    i\Big(\partial_t +\vec{v}_c\cdot\nabla+ \frac{1}{2}\Delta_{\rm C}\Big)\hat{\psi}= -\frac{1}{2}\Big(\nabla^2+\frac{\nabla{\rho_c}}{\rho_c} \cdot\nabla\Big)\hat{\psi} +g\rho_c(\hat{\psi}+\hat{\psi}^\dagger) -\Delta_{\rm E} \hat{\psi}.
\end{align}
Taking the real and imaginary parts of Eq.~\eqref{eq:BdGpsifluid} and using the continuity-like equation \eqref{eq:ClassCont}, we obtain Eqs.~\eqref{eq:QuantumFluid2} and \eqref{eq:QuantumFluid3}. Using the (formal) inverse of the operator $D$, we get the following wave equation:
\begin{align}
    -\frac{1}{\rho_c}\Big(\partial_t+\nabla\cdot\vec{v}_c-\frac{1}{2}\Delta_{\rm C}\Big)\rho_cD^{-1}\Big(\partial_t+\vec{v}_c\cdot\nabla+\frac{1}{2}\Delta_{\rm C}\Big)\delta\hat{\theta} + \Big(\frac{1}{\rho_c}\nabla\cdot(\rho_c\nabla)+2\Delta_{\rm E}\Big)\delta\hat{\theta} = 0.\label{eq:App:WaveEq}
\end{align}
This equation has the form of the wave equation of the linear theory with additional terms. With $c^2=D$ as in section~\ref{sec:sound}, Eq.~\eqref{eq:App:WaveEq} takes the form of Eq.~\eqref{eq:App:GeometricProto} with
\begin{align*}
    V &= -\frac{1}{2c^2}(\partial_t+\nabla\cdot\vec{v}_c)(\rho_c\Delta_{\rm C})+\frac{1}{4 c^2}\rho_c\Delta_{\rm C}^2  +2\rho_c\Delta_{\rm E}.
\end{align*}
Using the continuity-like equation \eqref{eq:ClassCont}, this reduces to the expression for $V$ given in section~\ref{sec:sound}. Finally, the identification $f^{\mu\nu} = \sqrt{-\mathfrak{g}}\,\mathfrak{g}^{\mu\nu}$ gives
\begin{equation*}
    \sqrt{-\mathfrak{g}} = \Big(\frac{\rho_c^d}{c^2}\Big)^{\frac{1}{d-2}},
\end{equation*}
from which the metric $\mathfrak{g}_{\mu\nu}$ and the mass $m^2$ of section~\ref{sec:sound} follow.

\section{Details of the model}\label{app:model}
In this appendix, we give the details of the calculation for the model of section~\ref{sec:model} \cite{Baak2022}. Let us denote the linear theory variables with subscript $0$, i.e.,
\begin{equation*}
    \hat{\Psi} = \phi_0(1 + \hat{\psi}_0),
\end{equation*}
which yields Gross-Pitaevskii equation
\begin{align}\label{eq:GPlinear}
    i\partial_t\phi_0 &= \Big(-\frac{1}{2}\nabla^2+U_{\rm ext} +g|\phi_0|^2\Big)\phi_0
\end{align}
and Bogoliubov-de Gennes equation:
\begin{align*}
    i\partial_t\hat{\psi}_0 &= \Big(-\frac{1}{2}\nabla^2-\frac{\nabla\phi_0}{\phi_0}\cdot\nabla\Big)\hat{\psi}_0 +g|\phi_0|^2(\hat{\psi}_0+\hat{\psi}_0^\dagger).
\end{align*}
For the potential of section~\ref{sec:model}, Eq.~\eqref{eq:GPlinear} has the stationary homogeneous solution
\begin{equation*}
    \phi_0 = \e^{-i\mu t}\sqrt{\rho_0}\qquad \textrm{ where } \qquad \rho_0 = |\phi_0|^2.
\end{equation*}
In here, $\rho_0$ is constant. For the quench from the noninteracting state,
\begin{equation*}
\hat{\psi}_0(t,x)=\sum_{n=0}^{\infty}\left[\hat{a}_{n}u_{n}^{(0)}(t,x)+\hat{a}^{\dagger}_{n}v_{n}^{(0)*}(t,x)\right],
\end{equation*}
where
\begin{align*}
    u_0^{(0)}(t,x) &= \frac{1}{\sqrt{\rho_0}}\frac{1}{\sqrt{\ell}} (1-it),\\
    v_0^{(0)}(t,x) &= \frac{1}{\sqrt{\rho_0}}\frac{1}{\sqrt{\ell}} (it),\\
    u_n^{(0)}(t,x) &= \frac{1}{\sqrt{\rho_0}}\frac{1}{\sqrt{2\ell}}\bigg[\cos{\omega_n t}-\frac{i}{\omega_n}\Big(\frac{k_n^2}{2}+1\Big)\sin{\omega_n t}\bigg](\e^{ik_n x}+(-1)^n\e^{-ik_n x}),\\
    v_n^{(0)}(t,x) &= \frac{1}{\sqrt{\rho_0}}\frac{i}{\sqrt{2\ell}}\frac{\sin{(\omega_n t)}}{\omega_n}\big(\e^{ik_n x}+(-1)^n\e^{-ik_n x}\big),
\end{align*}
$k_n=n\pi/\ell$, $\omega_n=\sqrt{k_n^2(k_n^2/4+1)}$, and $n=0,1,2,3,\ldots$.
Let us write
\begin{align*}
    \delta\hat{\rho}_0 \coloneq 2\rho_0\Re[\hat{\psi}_0],\qquad \delta\hat{\theta}_0 \coloneq \Im[\hat{\psi}_0].
\end{align*}
The correlation function at time $t$ of the sound wave in linear approximation is
\begin{align*}
    C_{\delta\theta_0}(x,y) &\coloneq\langle\delta\hat{\theta}_0(t,x)\delta\hat{\theta}_0(t,y)\rangle\nonumber \\
    &= -\frac{1}{4}\sum_n\Big[u_n^{(0)}(x)v_n^{(0)*}(y)+ v_n^{(0)}(x)u_n^{(0)*}(y)-v_n^{(0)}(x)v_n^{(0)*}(y)-u_n^{(0)}(x)u_n^{(0)*}(y)\Big],
\end{align*}
where we omit time $t$ in the mode function to make notation simple.
The canonical commutation relation $[\hat{\psi}_0(t,x),{\hat{\psi}_0^{\dagger}(t,y)}]=\frac{1}{\rho_0}\delta(x-y)$ shows that the correlation function can be separated into
\begin{equation*}
    C_{\delta\theta_0}(x,y)= \frac{1}{4\rho_0}\delta(x-y)+G_{\delta\theta_0}(x,y),
\end{equation*}
where $G_{\delta\theta_0}(x,y)$ is the normal ordered correlation function which is finite also when $x=y$.

Now, let us solve the system when there is a backreaction. Since the background is dynamical, we cannot use the familiar technique of choosing a preferred vacuum at some $t>0$ and computing expectation values by a Bogoliubov transformation. Instead, we tackle the problem differently by exploiting the linearity of the fluid-dynamical form of the Bogoliubov-de Gennes equation~\eqref{eq:BdGpsifluid}.

The first step is doing the mode expansion at $t=0$. Since the condensate is uniform before the interaction quench, we have the same initial condition at $t=0$ and can perform the mode expansion there, with $u_n(0,x) = u_{n}^{(0)}(0,x)$ and $v_{n}^{*}(0,x)=v_{n}^{(0)*}(0,x)$. Let us put the mode expansion into Eq.~\eqref{eq:BdGpsifluid}. Since the equation is linear, we can solve the equation at each $n$ separately. Moreover, if we define spinor $\Phi_n = (u_n,v_n)^\intercal$, Eq.~\eqref{eq:BdGpsifluid} can be written in the form
\begin{equation}\label{eq:BdGSpinor}
    i\sigma_3D_t \Phi_n = D_{\rm BdG}\Phi_n,
\end{equation}
where
\begin{equation*}
    D_t \coloneq\Big(\partial_t+\vec{v}_c\cdot\nabla+ \frac{1}{2}\Delta_{\rm C}\Big)
\end{equation*}
and
\begin{equation*}
    D_{\rm BdG} = \Big[-\frac{1}{2}\Big(\nabla^2+\frac{\nabla{\rho_c}}{\rho_c} \cdot\nabla\Big)-\Delta_{\rm E}\Big]\mathds{1}_{2} +g\rho_c\sigma_4.
\end{equation*}
Here, $\sigma_i$ ($i=1,2,3$) are the Pauli matrices and $\mathds{1}_2$ is the $2\times2$ identity matrix, and $\sigma_4\coloneq \mathds{1}+\sigma_1$.

We numerically solve Eq.~\eqref{eq:BdGSpinor} and Eq.~\eqref{eq:amendedGPpsi} to obtain each mode function. The correlation function is
\begin{align*}
    C_{\delta\theta}(t,x,y) = -\frac{1}{4}\sum_n\Big[u_n(x)v_n^*(y) + v_n(x)u_n^*(y)-v_n(x)v_n^*(y)-u_n(x)u_n^*(y)\Big],
\end{align*}
and the subtraction term of section~\ref{sec:model} is
\begin{align*}
    \Pi(x,y)= \sum_n\big[u_n(x)u_n^*(y) {- v_n^*(x)v_n(y)}\big].
\end{align*}
Note that at $t=0$, $\Pi(x,y) = \delta(x-y)/\rho_0$. The correlation function, subtraction term, and normal ordered correlation function become those of the linear theory when the mode functions become the linear theory ones, that is when $u_n,v_n \to u_n^{(0)},v_n^{(0)}$.

\section{Numerical method}\label{app:numerics}
\subsection{Coupled evolution}
We use units $\hbar=m=\xi_0=1$ and consider the homogeneous box $x\in[-\ell/2,\ell/2]$ with $\ell=40$, $\rho_0=150$, $g=1/150$, $\mu=g\rho_0=1$, and $N=\rho_0\ell=6000$. The initial state is the vacuum of the initial mode operators $\hat a_n$; no thermal sampling or stochastic trajectories are used. The condensate is initially uniform, $\phi_c(0,x)=\sqrt{\rho_0}$, and the modes are $u_n(0,x)=\varphi_n(x)/\sqrt{\rho_0}$, $v_n(0,x)=0$ with $\varphi_0=1/\sqrt{\ell}$, $\varphi_n=\big(e^{ik_nx}+(-1)^ne^{-ik_nx}\big)/\sqrt{2\ell}$ for $n\geq1$, and $k_n=n\pi/\ell$.

We integrate the coupled Eqs.~\eqref{eq:amendedGPpsi} and \eqref{eq:BdGpsi} directly in time. The mode expansion of $\hat\psi$ is truncated at $n_{\max}$, and the contractions entering the source terms are evaluated as $\langle\hat\psi^\dagger\hat\psi\rangle=\allowbreak\sum_{n=0}^{n_{\max}}|v_n|^2$ and $\langle\hat\psi^2\rangle=\sum_{n=0}^{n_{\max}}u_nv_n^*$. The $n=0$ channel has zero Bogoliubov frequency; in the linear theory its solution is $u_0=(1-it)/\sqrt{\rho_0\ell}$, $v_0=i t/\sqrt{\rho_0\ell}$, while in the backreacted calculation it is evolved by the same coupled equations and is included both in the contractions and in the correlation function. The Neumann walls are imposed through the basis, and the homogeneous phase $e^{-i\mu t}$ is removed in a rotating frame.

\subsection{Discretization}
The fields are represented by signed Fourier coefficients on a reflection-doubled ring with the Neumann parity $\tilde{f}_{-j}=(-1)^j\tilde{f}_j$, and the local products are evaluated with padded fast Fourier transforms. Time integration uses a fourth-order Lawson Runge-Kutta scheme in which the homogeneous Bogoliubov block is exponentiated exactly: for the spatial Fourier coefficient $q_j=j\pi/\ell$ the propagator is $e^{-iH_j\tau}=\cos(\Omega_j\tau)\mathds{1}-i\,\sin(\Omega_j\tau)H_j/\Omega_j$ with $\Omega_j=\sqrt{q_j^2(q_j^2/4+\mu)}$ and $H_j=\begin{pmatrix}q_j^2/2+\mu&\mu\\-\mu&-(q_j^2/2+\mu)\end{pmatrix}$. Only the self-consistent inhomogeneous residual is evaluated at the four stages, and at each stage the two equations and their contractions are evaluated from the same stage state, so that there is no time lag in the backreaction. The parameters are given in table~\ref{tab:num}.

\begin{table}[ht]
\caption{Numerical parameters.}
\label{tab:num}
\centering
\begin{tabular}{l|c}
\hline
quasiparticle cutoff $n_{\max}$ & $500$, $1000$ \\
spatial cutoff $K_x$ & $2500$ \\
FFT padding (doubled ring) & $16384$ \\
time interval, time step & $0\leq t\leq5$, $\Delta t=0.0025$ \\
\hline
\end{tabular}
\end{table}

\subsection{Observable: canonical-contact subtraction}
For numerical calculation, we need to truncate the mode sum at some finite number of modes, i.e., for some large number $n_{\rm max}$,
\begin{equation*}
    \sum_n\to\sum_{n=0}^{n_{\rm max}}.
\end{equation*}
The phase correlation function is
\begin{equation}
C_{\delta\theta,n_{\max}}(t,x,y)=\frac14\sum_{n=0}^{n_{\max}}\big[u_n(t,x)-v_n(t,x)\big]\big[u_n(t,y)-v_n(t,y)\big]^*.
\end{equation}
Its equal-point value contains a regulated $\delta(0)$ that depends on the cutoff. To remove it, we subtract the equal-time canonical commutator of the truncated mode expansion, reconstructed from the evolved modes,
\begin{equation}
\Pi_{n_{\max}}(t;x,y)=\sum_{n=0}^{n_{\max}}\big[u_n(t,x)u_n^*(t,y)-v_n^*(t,x)v_n(t,y)\big],
\end{equation}
which equals, for the homogeneous fixed-background solution, $\Pi_{n_{\max}}=\tfrac{1}{\rho_0}\sum_{n=0}^{n_{\max}}\varphi_n(x)\varphi_n(y)$, a truncated delta function. With backreaction the coefficient of the contact term changes with the condensate density, so that the initial projector cannot be used as the subtraction kernel at later times. The subtracted correlation is
\begin{equation}\label{eq:Gdeltathetanmax}
    G_{\delta\theta,n_{\max}}(t,x)=C_{\delta\theta,n_{\max}}(t,x,x)-\frac{1}{4}\Pi_{n_{\max}}(t;x,x)=\frac{1}{2}\sum_{n=0}^{n_{\max}}\Big[|v_n|^2-\mathrm{Re}(u_nv_n^*)\Big],
\end{equation}
evaluated mode by mode, and analogously for two points. The reference without backreaction is obtained with the linear modes,
\begin{equation}\label{eq:Gdeltatheta0nmax}
\rho_0G_{\delta\theta_0,n_{\max}}(t,x,y)=\frac{t^2}{\ell}+\sum_{n=1}^{n_{\max}}\frac{\sin^2(\omega_nt)}{k_n^2}\,\varphi_n(x)\varphi_n(y),\qquad \omega_n=\sqrt{k_n^2\Big(\frac{k_n^2}{4}+1\Big)},
\end{equation}
and the backreaction contribution is $\Delta G_{n_{\max}}=G_{\delta\theta,n_{\max}}-G_{\delta\theta_0,n_{\max}}$. We call this procedure canonical-contact subtraction (canonical normal ordering in section~\ref{sec:model}) and not a renormalization. The subtracted term is not a counterterm fixed by a renormalization condition, but the equal-time canonical commutator, a c-number determined by the operator algebra. Subtracting it removes precisely the $u_nu_n^*$ sum responsible for the short-distance (high-mode) divergence of $C_{\delta\theta}(t,x,x)$, since $v_n$ decays faster at high mode number. This is a structural cancellation, not a numerical coincidence: it holds mode by mode at every time, independently of the value of the remaining finite part.

\subsection{Convergence}
Table~\ref{tab:conv} compares two cutoffs at $t=5$ and $x=0$. The corresponding full-profile relative $L^2$ changes between $n_{\max}=500$ and $1000$ are $0.022\%$ for $G_{\delta\theta,n_{\max}}(5,x)$ and $0.39\%$ for $\Delta G_{n_{\max}}(5,x)$, comparable to the values at the center, which shows that the cutoff error is not concentrated near the walls. Halving the time step from $0.0025$ to $0.00125$ at $n_{\max}=500$ changes the center values of $G_{\delta\theta,n_{\max}}(5,0)$ and $\Delta G_{n_{\max}}(5,0)$ by $1.5\times10^{-7}$ and $1.5\times10^{-6}$ in relative terms, respectively; the larger relative change for $\Delta G_{n_{\max}}$ reflects its smaller magnitude ($\Delta G_{n_{\max}}\approx G_{\delta\theta,n_{\max}}/10$) rather than a larger absolute error. Both values are four to six orders of magnitude smaller than the cutoff dependence reported above.
\begin{table}[ht]
\caption{Cutoff dependence of the equal-point correlation at the center of the box. Values are given in the rescaled form $\rho_0G$, as in section~\ref{sec:results}.}
\label{tab:conv}
\centering
\begin{tabular}{l|ccc}
\hline
quantity at $t=5$, $x=0$ & $n_{\max}=500$ & $n_{\max}=1000$ & relative change \\
\hline
$G_{\delta\theta,n_{\max}}$ & $2.7639$ & $2.7645$ & $0.020\%$ \\
$G_{\delta\theta_0,n_{\max}}$ & $2.4885$ & $2.4905$ & $0.079\%$ \\
$\Delta G_{n_{\max}}$ & $0.2755$ & $0.2740$ & $0.52\%$ \\
\hline
\end{tabular}
\end{table}
Two further diagnostics were evaluated for $n_{\max}=1000$. First, since $\delta\hat\Psi=\phi_c\hat\psi$ obeys the canonical commutation relation, the contact coefficient reconstructed from the evolved modes should follow the instantaneous condensate density, $\Pi_{n_{\max}}/\Pi^{(0)}_{n_{\max}}-1\simeq\rho_0/\rho_c-1$, where $\rho_c=|\phi_c|^2$ and $\Pi^{(0)}_{n_{\max}}$ denotes the value without backreaction. The maximum deviation from this relation at $t=5$ is $1.1\times10^{-5}$, while $\rho_0/\rho_c-1$ itself reaches a few percent; the condensate and the modes, which are evolved by separate equations, thus remain mutually consistent, and the subtracted term is the local canonical contact term with its backreaction-modified coefficient. Second, we inspect the remaining high-mode contribution through the octave increment $D(M)=G_{\delta\theta,2M}(t,0)-G_{\delta\theta,M}(t,0)$ of the partial sums. Logarithmic growth of the partial sums with $M$ would give a constant $D(M)$; for $80\leq M\leq500$, however, a fit $a/M+b/M^2$ has a root-mean-square residual $6.8$ times smaller than a constant fit. This finite-range evidence is consistent with a summable inverse-cutoff tail, but it is not a proof of the asymptotic form.

\section{Hierarchical expansion and validity regime}\label{app:validity}
\begin{figure}[ht]
\centering
\begin{minipage}{0.62\textwidth}
\includegraphics[width=\linewidth]{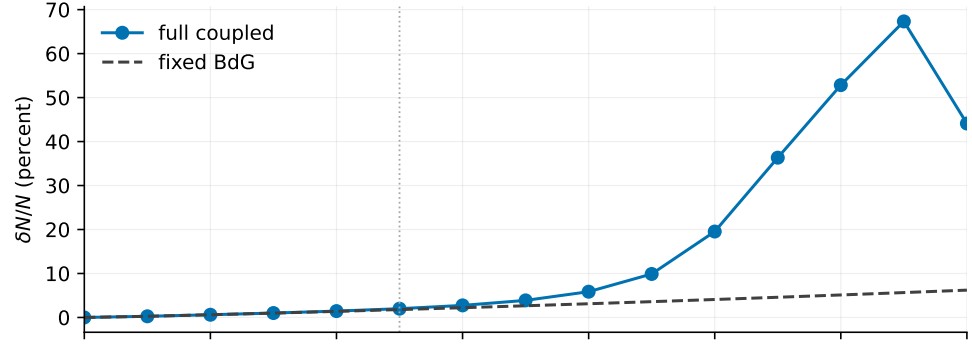}
\end{minipage}\hfill
\begin{minipage}{0.34\textwidth}
\includegraphics[width=\linewidth]{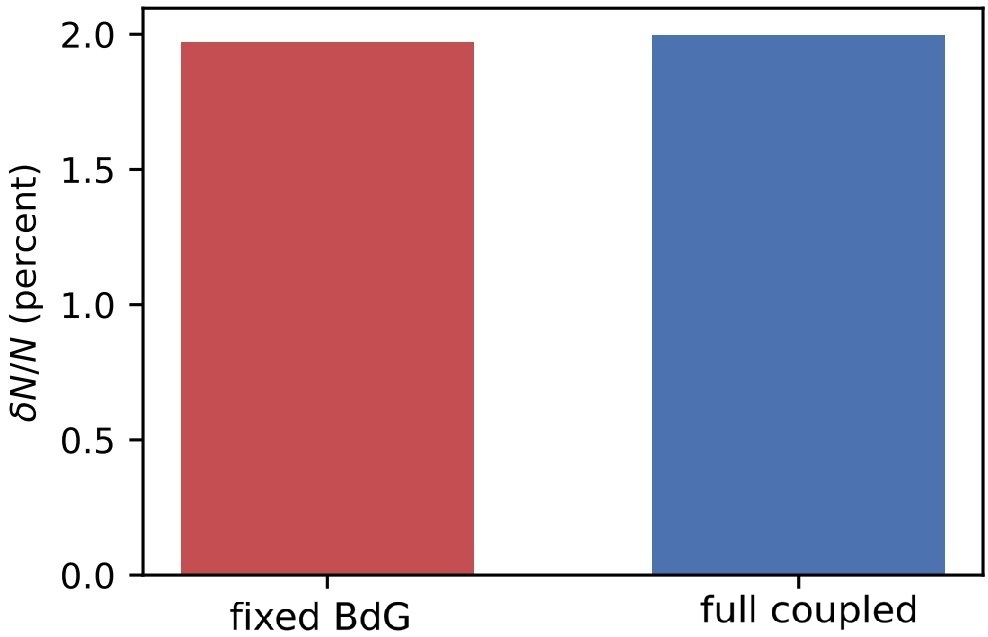}
\end{minipage}
\caption{Ratio $\delta N/N$ (in percent) of the number of depleted particles $\delta N$, i.e., the depletion integrated over space, to the total particle number $N$, for $\rho_0=150$ and $\ell=40$ at zero temperature. (Left) The solid line is the full (self-consistent) solution with backreaction, obtained by solving the coupled Eqs.~\eqref{eq:amendedGPpsi} and \eqref{eq:BdGpsi} numerically. The dashed line is the linear theory without backreaction, ``fixed BdG'', obtained by solving the Bogoliubov-de Gennes equation on the stationary classical field without backreaction. The two solutions agree closely up to $t\approx6$; the full solution then starts to grow faster, reaching about $10\%$ at $t\approx9$ and about $70\%$ at $t\approx13$, before decreasing. (Right) Values at $t=5$, the time used in section~\ref{sec:results}: the ratio is below $2\%$ for both solutions.}
\label{fig:Validity}
\end{figure}

\begin{figure}[ht]
\centering
\begin{minipage}{0.48\textwidth}
\includegraphics[width=\linewidth]{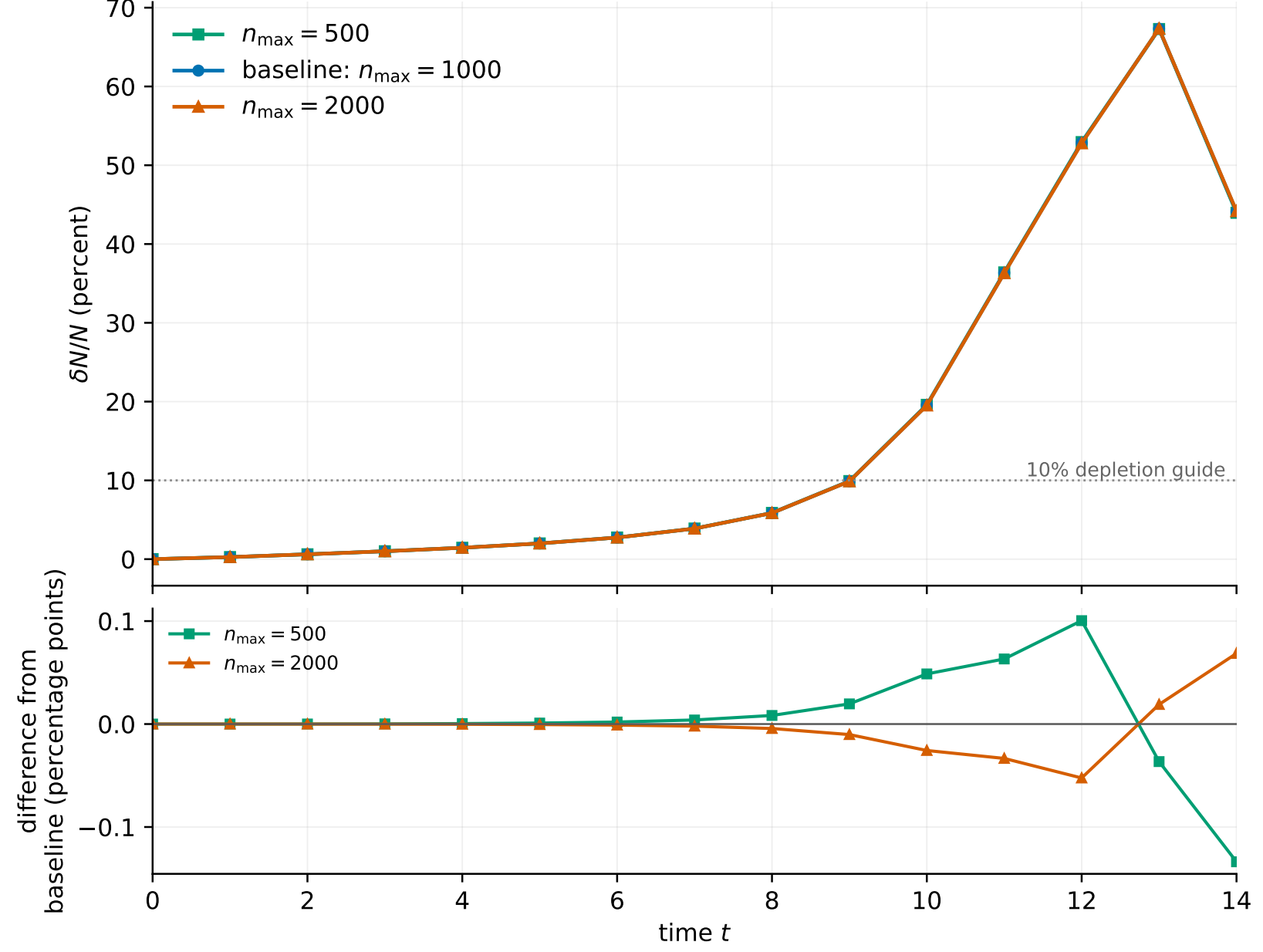}
\end{minipage}\hfill
\begin{minipage}{0.48\textwidth}
\includegraphics[width=\linewidth]{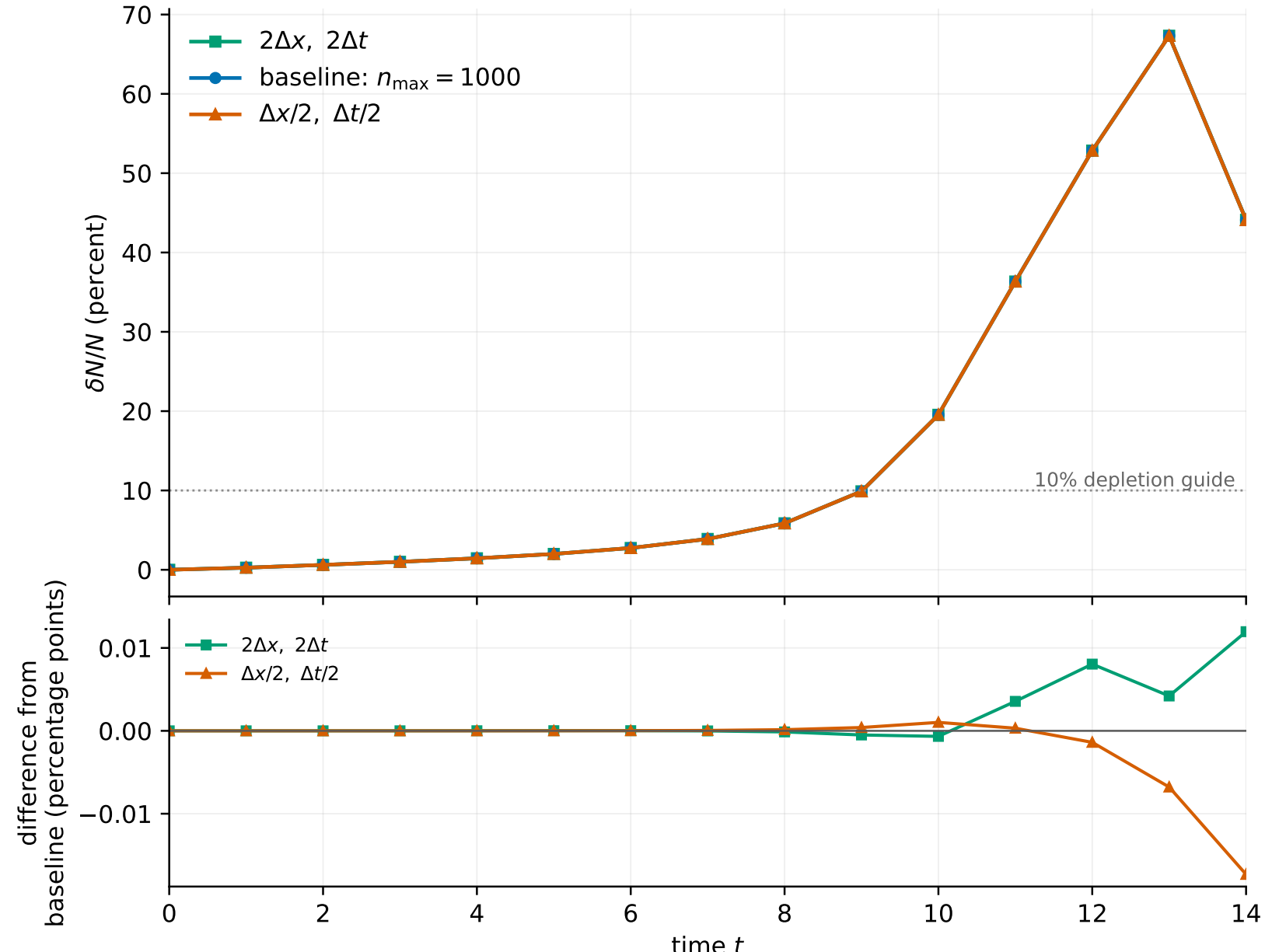}
\end{minipage}
\caption{Depletion $\delta N/N$ for the coupled solution, for different mode numbers $n_{\max}$, time steps $\Delta t$, and spatial resolutions $\Delta x$. The growth pattern starting at $t\approx6$ is unaffected by these numerical parameters over the range tested, indicating that it is not caused by numerical discretization.}
\label{fig:resolution_check}
\end{figure}

We have solved the problem by directly solving the coupled equations and evolving both $\phi_c$ and ${\hat{\psi}}$ together. For short times, one can also solve the problem by solving the hierarchy of uncoupled equations to obtain an approximate solution of $\phi_c$. That is, we can write the classical fluid observable, Eq.~\eqref{eq:classicalobservables} in the form of
\begin{equation*}
    \rho_c \simeq \rho_0+\rho_\zeta,\qquad \vec{\jmath}_c \simeq \vec{\jmath}_0+\vec{\jmath}_\zeta
\end{equation*}
with $\vec{\jmath}_0=0$ \cite{Baak2022}. Using the exact analytic form of $\rho_\zeta$ and $\vec{\jmath}_\zeta$, we can obtain each coefficient of Eq.~\eqref{eq:BdGpsifluid} explicitly
\begin{align*}
    \frac{\nabla\rho_c}{\rho_c}&\simeq \frac{\nabla\rho_\zeta}{\rho_0}= \frac{1}{2\rho_0\ell}\sum_{n=1}^{\infty}\frac{(-1)^n}{\omega_n^2}\sin(2k_n x)\left[\frac{2-k_n^2}{1+k_n^2}+2\cos(2\omega_n t)-\frac{k_n^2+4}{k_n^2+1}\cos(\omega_{2n}t)\right]\\
    v_c &\simeq \frac{j_\zeta}{\rho_0} = -\frac{2}{\rho_0\ell}\sum_{n=1}^\infty\frac{(-1)^n\sin(2k_nx)}{k_n}\left[\frac{\sin(2\omega_nt)}{2\omega_n}-\frac{\sin(\omega_{2n}t)}{\omega_{2n}}\right]
\end{align*}
\begin{align*}
    \Delta_{\rm C} &= 2g\rho_c\Im[\langle\hat{\psi}^2\rangle] \simeq 2g\rho_0\Im[\langle\hat{\psi}_0^2\rangle]\nonumber\\
    &= 2g\bigg[-\frac{t}{\ell} -\frac{1}{2\ell}\sum_{n=1}^{\infty} \frac{(-1)^n}{\omega_n}\big[(-1)^n+\cos(2k_nx)\big]\sin{(2\omega_n t)}\bigg],\\
    \Delta_{\rm E} &= g\rho_c\big(2\langle\hat{\psi}^\dagger\hat{\psi}\rangle + \Re[\langle\hat{\psi}^2\rangle]\big) \simeq g\rho_0\big(2\langle\hat{\psi}_0^\dagger\hat{\psi}_0\rangle + \Re[\langle\hat{\psi}_0^2\rangle]\big)\nonumber\\
    &= g\bigg[\frac{t^2}{\ell}+\frac{1}{2\ell}\sum_{n=1}^\infty\frac{(-1)^n}{\omega_n^2}\Big(1-\frac{k_n^2}{2}\Big)\big[(-1)^n+\cos(2k_nx)\big]\big[1-\cos(2\omega_nt)\big]\bigg].
\end{align*}
Here, the zero-mode contribution is also taken into account, since in the hierarchical expansion the linear field is obtained in the Bogoliubov approximation \cite{Baak2024,Baak2022}. Then, we can solve linear equation Eq.~\eqref{eq:BdGpsifluid} directly using the same method. Note that the number-conserving expansion is valid in the regime where the depletion is much smaller than the total number of particles. We monitor the ratio $\delta N/N$ of the number of depleted particles to the total particle number in figure~\ref{fig:Validity}. Since the difference between the coupled and the linear-theory solutions is, by construction, the leading $O(1/N)$ correction of the expansion, its smallness is itself a direct indicator of the validity of the approximation. For the parameters used in section~\ref{sec:results}, $\delta N/N$ is below $2\%$ up to $t=5$ in both methods, which agree closely in this regime. When $t\gtrsim6$, the depletion for the coupled numerical solution starts to grow faster than the linear one; by $t\approx9$ it has grown to about $10\%$, and for larger $t$ it grows rapidly, so that our approximation is valid only in a smaller window. This growth is unaffected by the mode number, the time step, and the spatial resolution over the range tested (figure~\ref{fig:resolution_check}), indicating that it is not caused by numerical discretization. This window size depends on the number of particles.

Within this scheme, the backreaction terms in Eq.~\eqref{eq:BdGpsi} shift the chemical potential of the fluctuations, whereas their Hartree-Fock interaction with the depleted fraction, which enters at the same order through terms cubic in $\hat{\psi}$, is not included. As a consequence, the spectrum of a stationary depleted condensate is not gapless at this order: for a homogeneous condensate, Eqs.~\eqref{eq:QuantumFluid2} and \eqref{eq:QuantumFluid3} give $\omega^2\simeq c^2(k^2-2\Delta_{\rm E})$ at long wavelengths, corresponding to a negative mass squared. This is the gap problem known from the Hartree-Fock-Bogoliubov theory of Bose gases; gapless treatments, such as the Popov approximation, which includes this interaction and neglects the anomalous average \cite{Griffin1996,Proukakis2008}, would modify this term. We therefore restrict our numerical results to $t\leq5$, where the depletion is small, and note that the rapid growth of the depletion at later times may be related to this long-wavelength instability.
\begin{table}[ht]
\caption{Depletion at $t=5$ for several particle numbers ($N=40\rho_0$). The linear-theory depletion scales as $1/N$, while the relative difference between the coupled and linear solutions, the leading $O(1/N)$ correction to the depletion itself, grows as $\rho_0$ decreases, signaling the earlier onset of the breakdown of the expansion for smaller $N$.}
\label{tab:depletion_scan}
\centering
\begin{tabular}{c|cccc}
\hline
$\rho_0$ & 50 & 100 & 300 & 450 \\
\hline
linear ($t=5$) & $5.27\%$ & $2.69\%$ & $0.89\%$ & $0.597\%$ \\
coupled ($t=5$) & $7.45\%$ & $3.16\%$ & $0.946\%$ & $0.619\%$ \\
relative difference & $41\%$ & $17\%$ & $6.3\%$ & $3.7\%$ \\
\hline
\end{tabular}
\end{table}
\begin{figure}[ht]
\centering
\includegraphics[width=0.8\linewidth]{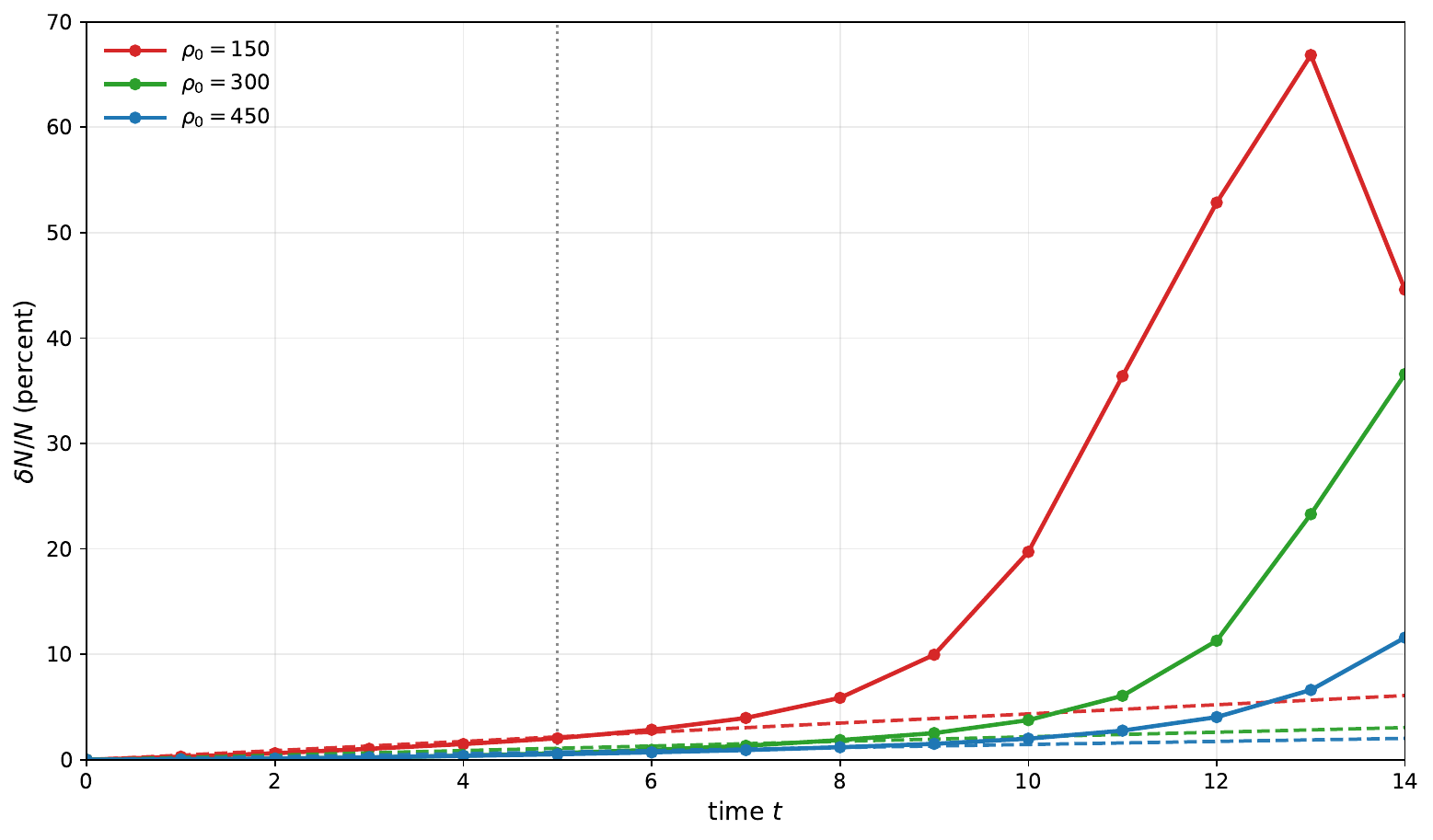}
\caption{Depletion $\delta N/N$ as a function of time, for $\rho_0=150,300,450$. Solid lines are the coupled solution; dashed lines are the linear theory. The onset of rapid growth moves to later times as $\rho_0$ increases.}
\label{fig:depletion_scan}
\end{figure}

\begin{figure}[ht]
\centering
\begin{minipage}{0.48\textwidth}
\includegraphics[width=\linewidth]{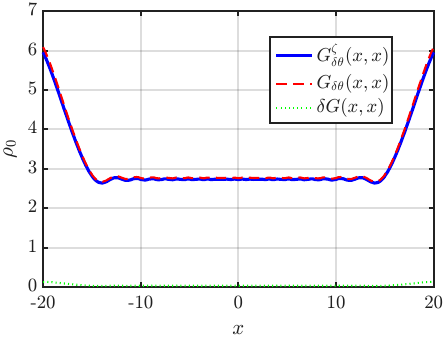}
\end{minipage}\hfill
\begin{minipage}{0.48\textwidth}
\includegraphics[width=\linewidth]{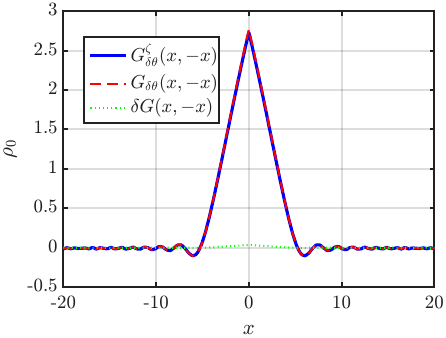}
\end{minipage}
\caption{Canonically normal-ordered correlation function at the same (left panel) and opposite (right panel) points at $t=5$. The red dashed line $G_{\delta\theta}$ is obtained from the self-consistent evolution of the coupled equations used in section~\ref{sec:results}; the blue line $G_{\delta\theta}^\zeta$ is obtained from the hierarchical expansion method described above; the green dotted line shows their difference $\delta G$. The difference is about $2\%$ of the correlation function itself at the same point and about $1\%$ at opposite points.}
\label{fig:Check}
\end{figure}

As a further, independent check, we verify that the self-consistent treatment of backreaction used in section~\ref{sec:results} agrees with the hierarchical expansion described above, in which the backreaction of the quantum field on the classical background is computed first, and the resulting correction to the quantum field is then computed on this backreacted background. In figure~\ref{fig:Check}, we plot $G_{\delta\theta}$ (coupled) and $G_{\delta\theta}^\zeta$ (hierarchical), together with their difference $\delta G\coloneqq G_{\delta\theta}-G_{\delta\theta}^\zeta$. As the figure shows, the difference is small, confirming that both approaches agree closely at $t=5$.

\section{Effect of finite imaging resolution}\label{app:imaging}
\subsection{Mapping to physical units}
All numerical results refer to $\rho_0\xi_0=150$, $\ell=40\,\xi_0$, and $t=5\,m\xi_0^2/\hbar$ at zero temperature, with the correlation functions computed as described in appendix~\ref{app:numerics} ($n_{\max}=1000$). The linear-theory reference $G^{(0)}$ used below is evaluated from its analytic mode sum, which agrees with the numerically computed linear solution to $10^{-13}$, and the cutoff dependence of the backreaction contribution is below $0.6\%$ (table~\ref{tab:conv}). Because the results depend only on these dimensionless combinations, physical units follow from a choice of the atomic species and of $\xi_0$: the linear density per gas is $n=\rho_0=150/\xi_0$, the sound velocity is $c=\hbar/(m\xi_0)$, and the unit of time is $m\xi_0^2/\hbar$, where $m$ is the atomic mass. Here $g$ is the coupling constant of the one-dimensional model, with the dimension of energy times length, fixed by $g\rho_0=mc^2=\hbar^2/(m\xi_0^2)$, i.e., $g=1/(\rho_0\xi_0)$ in units of $\hbar^2/(m\xi_0)$. For a gas with tight transverse confinement of frequency $\omega_\perp$ it is related to the three-dimensional s-wave scattering length $a_s$ by $g\simeq2\hbar\omega_\perp a_s$ \cite{Olshanii1998}, so that
\begin{equation}
a_s=\frac{\hbar}{2m\omega_\perp\xi_0^2\rho_0},\qquad
\frac{\mu}{\hbar\omega_\perp}=\frac{g\rho_0}{\hbar\omega_\perp}=2a_s\rho_0 .
\label{eq:as}
\end{equation}
The second relation measures how well the one-dimensional condition $\mu\ll\hbar\omega_\perp$ is satisfied. Table~\ref{tab:map} lists three boxes for $^{87}$Rb ($\hbar/m=0.731\,\mu\mathrm{m}^2/\mathrm{ms}$) and $\omega_\perp=2\pi\times1.4$ kHz. The box of $13.2\,\mu$m is shorter than those of present experiments and, with $\mu\approx0.76\,\hbar\omega_\perp$, only marginally one-dimensional; it is included for comparison only.
\begin{table}[ht]
\caption{Mapping of the dimensionless calculation, $\rho_0\xi_0=150$ and $\ell=40\,\xi_0$, to three box lengths for $^{87}$Rb with $\omega_\perp=2\pi\times1.4$ kHz.}
\label{tab:map}
\centering
\begin{tabular}{l|ccc}
\hline
box length $\ell$ ($\mu$m) & 13.2 & 46 & 100 \\
$\xi_0$ ($\mu$m) & 0.33 & 1.15 & 2.5 \\
density per gas $n$ ($\mu\mathrm{m}^{-1}$) & 455 & 130 & 60 \\
sound velocity ($\mu$m/ms) & 2.2 & 0.64 & 0.29 \\
unit of time (ms) / $t=5$ (ms) & 0.15 / 0.75 & 1.8 / 9.0 & 8.6 / 43 \\
required $a_s$ (nm) & 0.84 & 0.24 & 0.11 \\
$\mu/(\hbar\omega_\perp)$ & 0.76 & 0.063 & 0.013 \\
number of $2\,\mu$m pixels & 6 & 23 & 51 \\
pixel / PSF ($3\,\mu$m) in units of $\xi_0$ & 6.1 / 9.1 & 1.7 / 2.6 & 0.78 / 1.2 \\
\hline
\end{tabular}
\end{table}
The required scattering lengths in table~\ref{tab:map} are considerably smaller than the natural value for $^{87}$Rb, $a_s=5.2$ nm. From Eq.~\eqref{eq:as}, the dimensionless parameter of the calculation is
\begin{equation}
\rho_0\xi_0=\sqrt{\frac{\hbar\,n}{2m\omega_\perp a_s}}
=\frac{n\,a_\perp}{\sqrt{\mu/(\hbar\omega_\perp)}},\qquad a_\perp=\sqrt{\frac{\hbar}{m\omega_\perp}},
\label{eq:rhoxi}
\end{equation}
which depends on the density and the interaction strength but not on the box length. At the natural scattering length and the linear densities realized in present experiments, $n\approx37\ \mu\mathrm{m}^{-1}$ per gas in Ref.~\cite{Murtadho2025} and $n\approx100\ \mu\mathrm{m}^{-1}$ per gas in Ref.~\cite{Tajik2023}, this gives $\rho_0\xi_0\approx17$ and $28$, respectively, well below the value $\rho_0\xi_0=150$ used here (table~\ref{tab:natural}); a longer box at the same density and scattering length does not change this value. Reaching $\rho_0\xi_0=150$ at the natural scattering length and $\omega_\perp=2\pi\times1.4$ kHz would instead require $n\approx2.8\times10^3\ \mu\mathrm{m}^{-1}$, for which $\mu\approx29\,\hbar\omega_\perp$, far outside the one-dimensional regime. Weakening the transverse confinement does not help either: at the natural scattering length, $\rho_0\xi_0=150$ with $\mu\approx0.1\,\hbar\omega_\perp$ would require $n\approx10\ \mu\mathrm{m}^{-1}$ and $\omega_\perp\approx2\pi\times5$ Hz. The regime studied here therefore requires tuning the interaction well below its natural value, for example through a Feshbach resonance \cite{Chin2010}. Table~\ref{tab:natural} also lists, for each density, the scattering length and the box length for which $\rho_0\xi_0=150$ and $\ell=40\,\xi_0$; the density of the $46\,\mu$m box of table~\ref{tab:map}, $n\approx130\ \mu\mathrm{m}^{-1}$, is included for comparison.
\begin{table}[ht]
\caption{Healing length, $\rho_0\xi_0$, and $\mu/(\hbar\omega_\perp)$ for $^{87}$Rb with $\omega_\perp=2\pi\times1.4$ kHz at its natural scattering length (upper part), and the scattering length and box length required for $\rho_0\xi_0=150$ and $\ell=40\,\xi_0$ (lower part), at the densities of Refs.~\cite{Murtadho2025} and \cite{Tajik2023} and of the $46\,\mu$m box in table~\ref{tab:map}. For $\mu\gtrsim\hbar\omega_\perp$ the one-dimensional relations used here are only indicative.}
\label{tab:natural}
\centering
\begin{tabular}{l|ccc}
\hline
density per gas $n$ ($\mu\mathrm{m}^{-1}$) & 37 & 100 & 130 \\
\hline
\multicolumn{4}{l}{natural scattering length, $a_s=5.2$ nm} \\
\quad $\xi_0$ ($\mu$m) & 0.47 & 0.28 & 0.25 \\
\quad $\rho_0\xi_0$ & 17 & 28 & 32 \\
\quad $\mu/(\hbar\omega_\perp)$ & 0.38 & 1.0 & 1.4 \\
\hline
\multicolumn{4}{l}{tuned to $\rho_0\xi_0=150$, $\ell=40\,\xi_0$} \\
\quad $\xi_0$ ($\mu$m) & 4.1 & 1.5 & 1.15 \\
\quad box length $\ell$ ($\mu$m) & 162 & 60 & 46 \\
\quad required $a_s$ (nm) & 0.068 & 0.18 & 0.24 \\
\quad $\mu/(\hbar\omega_\perp)$ & 0.005 & 0.037 & 0.063 \\
\hline
\end{tabular}
\end{table}
The box of $46\,\mu$m in table~\ref{tab:map} matches the length realized in Ref.~\cite{Murtadho2025}, and the required density, $n\approx130\,\mu\mathrm{m}^{-1}$, is comparable to that reached in Ref.~\cite{Tajik2023}. The box length and density are thus individually close to values already realized in present experiments, whereas the required scattering length, $a_s\approx0.24$ nm, about $20$ times smaller than the natural value, has not been combined with these setups. Since the Feshbach resonances of $^{87}$Rb are narrow \cite{Chin2010}, such a reduction requires precise control of the magnetic field; species with broader resonances would make it less demanding, at the cost of different numbers in table~\ref{tab:map}. The parameters $\rho_0\xi_0=150$ and $\ell=40\,\xi_0$ used in the calculation therefore correspond to a physically realistic combination, although one that has not yet been demonstrated.

\subsection{Imaging model}
We model finite resolution as in Refs.~\cite{Murtadho2025,Moller2026}: the phase field is convolved with a Gaussian of width $\sigma$ and then averaged over the camera pixels of $2\,\mu$m, the pixel size of vertical imaging in Refs.~\cite{Tajik2023,Murtadho2025}. Both operations are linear, so the phase correlation function $G(x,y)$ is transformed into $FGF^{T}$, where $F$ is the product of the Gaussian filter and the pixel averaging. The filter uses reflecting boundaries because the mode functions satisfy Neumann conditions at the walls, and grid points at the ends that do not fill a complete pixel are discarded symmetrically. For two independent identical gases the correlation of the relative phase is twice that of a single gas; this factor cancels in relative differences. The imaging process of real experiments comprises absorption of the probe light, recoil blurring, an optical transfer function that depends on the cloud size and defocus, pixel binning, and shot noise, and the effective width of the point spread function is usually fitted \cite{Moller2021}; the Gaussian model is therefore a simplification.

Since the global phase cannot be measured, the reconstruction of Ref.~\cite{Murtadho2025} fixes it by $\int\phi\,dz=0$ and is restricted to the bulk of the gas. We mimic this by keeping the central $80\%$ of the pixels and subtracting the mean of the phase over these pixels, i.e., by applying the centering matrix $P=\mathds{1}-\mathbf{1}\mathbf{1}^T/n_{\mathrm{p}}$ to both indices of the correlation matrix, where $\mathds{1}$ is the identity, $\mathbf{1}=(1,\dots,1)^T$, and $n_{\mathrm{p}}$ is the number of retained pixels. This removes the $x$-independent contribution of the zero mode.

\subsection{Same-point phase correlation}
Table~\ref{tab:sigma} shows the relative increase $\Delta G_{n_{\max}}/G_{\delta\theta_0,n_{\max}}$ of the same-point phase correlation as a function of the width $\sigma$ of the Gaussian, for the box of $46\,\mu$m. ``Outer'' is the average over the outermost camera pixel at both ends, and ``bulk'' is the average over the central $80\%$ of the pixels after removing the global mode. For the boxes of $13.2\,\mu$m and $100\,\mu$m the bulk values lie between $12.5\%$ and $12.8\%$ and between $11.4\%$ and $12.3\%$, respectively, for all $\sigma\leq9\,\mu$m, while the outermost pixel gives $12$ to $13.5\%$ and $17$ to $22.5\%$. For $\sigma\gtrsim\ell/4$ the filter averages over a large part of the box and the results lose their meaning.

\begin{table}[ht]
\caption{Relative increase of the same-point phase correlation for the box of $46\,\mu$m as a function of the width of the Gaussian.}
\label{tab:sigma}
\centering
\begin{tabular}{l|cccccccc}
\hline
$\sigma$ ($\mu$m) & 0 & 0.5 & 1 & 2 & 3 & 4 & 6 & 9 \\
outer (\%) & 21.7 & 21.6 & 21.2 & 19.9 & 18.4 & 17.1 & 15.2 & 13.7 \\
bulk, global mode removed (\%) & 11.7 & 11.9 & 12.1 & 12.3 & 12.3 & 12.2 & 12.3 & 12.5 \\
\hline
\end{tabular}
\end{table}

Without imaging, the relative difference of the same-point correlation near the left wall is given in table~\ref{tab:wall}. It is $21.8\%$ at the wall, has a maximum of $22.3\%$ about $0.5\,\xi_0$ inside, and reaches the bulk value within about six healing lengths. The distances in micrometers refer to the box of $46\,\mu$m.

\begin{table}[ht]
\caption{Relative difference of the same-point phase correlation near the left wall, without imaging.}
\label{tab:wall}
\centering
\begin{tabular}{c|cccccccc}
\hline
distance from wall ($\xi_0$) & 0 & 0.5 & 0.9 & 2.0 & 3.0 & 5.9 & 10 & 20 \\
distance ($\mu$m) & 0 & 0.54 & 1.08 & 2.34 & 3.41 & 6.83 & 11.5 & 23.0 \\
$\Delta G_{n_{\max}}/G_{\delta\theta_0,n_{\max}}$ ($\%$) & 21.8 & 22.3 & 21.8 & 18.2 & 14.2 & 10.9 & 10.9 & 11.0 \\
\hline
\end{tabular}
\end{table}

\subsection{Velocity correlations}
Following the definition of the measured velocity field in Ref.~\cite{Tajik2023}, the measured phase at pixel $i$ is obtained by applying a Gaussian filter of width $\sigma$ followed by pixel averaging over a pixel of size $\Delta$. The velocity of pixel $i$ is then defined as the finite difference of the filtered phase $\overline{\delta\theta}_i$ between neighboring pixels,
\begin{equation}
v_i = \frac{\overline{\delta\theta}_{i+1} - \overline{\delta\theta}_i}{\Delta}.
\end{equation}
The velocity correlation between pixels $i$ and $j$ is therefore
\begin{align}
\langle v_iv_j\rangle &= \frac{1}{\Delta^2}\langle(\overline{\delta\theta}_{i+1}-\overline{\delta\theta}_i)(\overline{\delta\theta}_{j+1}-\overline{\delta\theta}_j)\rangle \notag\\
&=\frac{\overline{G}_{\delta\theta}(x_{i+1},x_{j+1})-\overline{G}_{\delta\theta}(x_{i+1},x_j)-\overline{G}_{\delta\theta}(x_i,x_{j+1})+\overline{G}_{\delta\theta}(x_i,x_j)}{\Delta^2},
\end{align}
where $\overline{G}_{\delta\theta}(x_i,x_j)\equiv\langle\overline{\delta\theta}_i\overline{\delta\theta}_j\rangle$ is the filtered phase correlation function, averaged over all pairs with fixed $|i-j|$. Table~\ref{tab:vel} compares the backreaction contribution $\Delta\langle v_iv_j\rangle$ with the linear-theory value $\langle v_iv_j\rangle_0$ for the box of $46\,\mu$m and $\sigma=3\,\mu$m. For neighboring pixels the relative increase is $12.8\%$ ($13.2\,\mu$m box), $16.1\%$ ($46\,\mu$m box), and $14.3\%$ ($100\,\mu$m box). Without the Gaussian, only the pixel averaging acts, and the ratios vary strongly with the pixel separation (from about $5\%$ to more than $200\%$) and are not meaningful where $\langle v_iv_j\rangle_0$ is small; we do not use them.
\begin{table}[ht]
\caption{Velocity correlations for the box of $46\,\mu$m and $\sigma=3\,\mu$m. Values are in units of $\rho_0^{-1}\xi_0^{-2}$.}
\label{tab:vel}
\centering
\begin{tabular}{c|ccc}
\hline
$|i-j|$ & $\Delta\langle v_iv_j\rangle$ & $\langle v_iv_j\rangle_0$ & ratio \\
\hline
0 & $8.7\times10^{-3}$ & $5.3\times10^{-2}$ & $16.4\%$ \\
1 & $7.9\times10^{-3}$ & $4.9\times10^{-2}$ & $16.1\%$ \\
2 & $5.2\times10^{-3}$ & $3.4\times10^{-2}$ & $15.3\%$ \\
3 & $1.9\times10^{-3}$ & $1.5\times10^{-2}$ & $12.1\%$ \\
\hline
\end{tabular}
\end{table}
For the box of $13.2\,\mu$m we tested the dependence on the alignment of the pixel grid with respect to the box: for $\sigma\geq1.5\,\mu$m the variation between eight alignments is below $3\%$ of the mean.

\subsection{Limitations}
The above results are for canonically normal-ordered correlation functions at zero temperature, whereas measured correlations contain the subtracted term and classical thermal phase fluctuations, whose magnitude relative to the backreaction contribution is not quantified here. The reported percentages are therefore approximate mainly because of these physical simplifications and the simplified imaging model, rather than because of numerical uncertainty: the cutoff dependence of the backreaction contribution is below $0.6\%$.

\subsection{Outlook}\label{app:outlook}
We briefly comment on how future experiments could improve access to this effect, combining three directions: the particle number, the interaction strength, and the temperature.

First, the window of validity of the expansion is set by the depletion, which at fixed $\ell/\xi_0$ scales as $1/N$, as expected for the $1/N$ expansion; for $\ell=40\,\xi_0$ and $t=5$, the linear-theory depletion is $\delta N/N\approx2.7/(\rho_0\xi_0)=2.7\,\ell/(\xi_0N)$. A larger particle number at the same box length and healing length would therefore extend the window. Since $\xi_0^2=\hbar/(2m\omega_\perp a_s n)$, this requires increasing the density and reducing the scattering length in proportion, $a_s\propto1/n$, which leaves $\mu/(\hbar\omega_\perp)$, the sound velocity, and the unit of time unchanged. For the box of $46\,\mu$m in table~\ref{tab:map}, doubling $N$ to about $1.2\times10^4$ requires $n\approx260\,\mu\mathrm{m}^{-1}$ and $a_s\approx0.12$ nm. Increasing $N$ instead by lengthening the box at fixed density and scattering length would not reduce the depletion appreciably, because the number of phonon modes that contribute to it grows in proportion to the box length.

Second, reaching the specific values of $\rho_0\xi_0$ and $\ell$ studied here requires tuning the interaction well below its natural value for $^{87}$Rb, as discussed above (tables~\ref{tab:map} and \ref{tab:natural}); for the box of $46\,\mu$m already realized in Ref.~\cite{Murtadho2025}, this means increasing the density to $n\approx130\,\mu\mathrm{m}^{-1}$, comparable to that reached in Ref.~\cite{Tajik2023}, and reducing the scattering length to $a_s\approx0.24$ nm via a Feshbach resonance \cite{Chin2010}, a reduction by a factor of about $20$ from the natural value.

Third, lower temperatures than those realized in Ref.~\cite{Murtadho2025} ($T\approx31$ nK) appear attainable: continuous, homogeneous outcoupling of atoms from a one-dimensional $^{87}$Rb quasicondensate on an atom chip has been shown to cool the gas, via dephasing of the phonon modes rather than conventional evaporative cooling, down to $T\approx20$ nK at a particle number $N\approx6000$, equal to $N=\rho_0\ell$ for the $46\,\mu$m box of table~\ref{tab:map}, starting from initial temperatures of $30$ to $100$ nK \cite{Rauer2016}. For a homogeneous system the same mechanism is predicted to follow $T\propto\rho_0^{3/2}$, as opposed to the linear scaling $T\propto N$ found for a harmonically trapped gas, so that a box trap could in principle reach comparably low temperatures with a smaller reduction in particle number, although this remains to be demonstrated experimentally.

Combining these directions is a natural, though as yet untested, target for future experiments: the box of $46\,\mu$m already realized in Ref.~\cite{Murtadho2025}, at the density and scattering length discussed above, cooled through the mechanism of Ref.~\cite{Rauer2016}. Each ingredient has been demonstrated separately; combining them, in particular adapting the cooling mechanism of Ref.~\cite{Rauer2016} from a harmonic to a box trap, remains an open experimental question.
\end{appendix}

\bibliography{BRcorr_scipost}

\end{document}